\documentclass[draft,numberedheadings]{aipproc}
 \layoutstyle{6x9}
  \usepackage{amsmath}
   \usepackage{amssymb}
    \usepackage{pifont}

\begin{document}

\title{Pion distribution amplitude -- from theory to data}


\author{A.~P.~Bakulev}
 {address={Bogoliubov Laboratory of Theoretical Physics,
           JINR, 141980 Dubna, Russia}\footnote{%
 Talk based on results obtained in collaboration
 with S.~Mikhailov, K.~Passek-Kumeri\v{c}ki, W.~Schroers, and
 N.~G.~Stefanis}}

\begin{abstract}
 We describe the present status of the pion distribution amplitude
 as originated from two sources:
 (i) a nonperturbative approach, based on QCD sum rules with nonlocal
 condensates, and
 (ii) a NLO QCD analysis of the CLEO data on
 $F^{\gamma\gamma^*\pi}(Q^2)$, supplemented by the E791 data on
 diffractive dijet production, and the JLab F(pi) data on the pion
 electromagnetic form factor.
\end{abstract}

\maketitle

\section{Pion distribution amplitude from QCD sum rules}
The pion distribution amplitude (DA)
can be defined through the matrix element
of a nonlocal axial current on the light cone
\begin{eqnarray}
 \langle{0\mid \bar d(z)\gamma_{\mu}\gamma_5
 {\cal C}(z,0) u(0)\mid \pi(P)}\rangle\Big|_{z^2=0}
 &=& i f_{\pi}P_{\mu}
    \int_{0}^{1} dx\ e^{ix(zP)}\
     \varphi_{\pi}(x,\mu^2)\,,
    \label{eq:PiME}
\end{eqnarray}
which is explicitly  gauge-invariant
due to the connector
${\cal C}(z,0)={\cal P}e^{i g \int_0^z A_\mu(\tau) d\tau^\mu}$.
This amplitude describes the transition of the physical pion $\pi(P)$
to a pair of valence quarks $u$ and $d$, separated on the light-cone,
with corresponding momentum fractions $xP$ and $\bar{x}P$,
(we set $\bar{x}\equiv 1-x$).

In order to obtain the pion DA theoretically we use,
following Mikhailov and Radyushkin~\cite{MR86}, a
QCD sum rule approach with non-local condensates (NLC).
Just for illustration, we present here the simplest scalar condensate
of the used NLC model, which reads
\begin{equation}
 \label{eq:ScaNLC}
  \langle{\bar{q}(0)q(z)}\rangle
   =\langle{\bar{q}(0)q(0)}\rangle\, e^{-|z^2|\lambda_q^2/8}\,.
\end{equation}
This model is determined by a single scale parameter $\lambda_q^2 = \langle{k^2}\rangle$
characterizing the average momentum of quarks in the QCD vacuum.
It has been estimated in QCD SRs~\cite{BI82,OPiv88} and on the lattice~\cite{DDM99,BM02}:
$\lambda_q^2 = 0.45\pm 0.1\text{~GeV}^2$.

\begin{figure}[t]
 $$\includegraphics[width=0.9\textwidth]{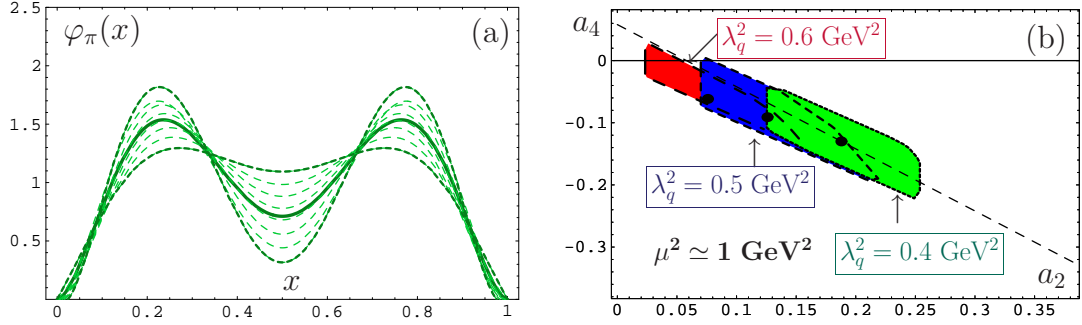}$$\vspace*{-9mm}
   \caption{\label{fig:b456}\footnotesize
    (a) The bunch of pion DAs extracted from NLC QCD sum rules.
     The parameters of the bold-faced curve are $a_2^\textbf{b.f.}=+ 0.188$
     and $a_4^\textbf{b.f.}= - 0.130$.
    (b) The ``bunches'' of pion DAs extracted from NLC QCD sum rules
    in the $(a_2,a_4)$-plane for three values of the nonlocality parameter $\lambda_q^2$.}
\end{figure}

The NLC sum rules for the pion DA produce~\cite{BMS01}
a ``\textbf{bunch}'' of self-consistent 2-parameter models at $\mu^2\simeq 1$ GeV$^2$:
\begin{equation}
 \label{eq:2p-Bunch}
  \varphi_\pi(x)
  = \varphi^{\text{as}}(x)
      \left[1 + a_2 C^{3/2}_2(2x-1) + a_4 C^{3/2}_4(2x-1)
      \right]\,.
\end{equation}
By self-consistency we mean that the value of the inverse moment
for the whole ``bunch''
$\langle{x^{-1}}\rangle_\pi^{\text{bunch}} = 3.17\pm0.10$
is in agreement with an independent estimate from another sum rule, viz.,
$\langle{x^{-1}}\rangle_\pi^{\text{SR}}=3.30\pm0.30$.
For the favored value $\lambda_q^2=0.4$ GeV$^2$,
we get the ``bunch'' of pion DAs presented in Fig.\,\ref{fig:b456}a.
We also extracted the corresponding ``bunches'' for two other values
of $\lambda_q^2=0.5$~GeV$^2$ and $\lambda_q^2=0.6$~GeV$^2$,
and show the results (rectangle areas) in the $(a_2,a_4)$-plane
in Fig.\,\ref{fig:b456}b.

\section{NLO light-cone\ sum rules (LCSR) and the CLEO data}
The CLEO experimental data on $F^{\gamma\gamma^*\pi}(Q^2)$ allow one
to obtain direct constraints on $\varphi_\pi(x)$.
For $Q^2\gg m_\rho^2$, $q^2\ll m_\rho^2$
pQCD factorization is valid only in leading twist,
but higher twists are also of importance~\cite{RR96}.
The reason is: if $q^2\to 0$ one needs to take into account
the interaction of a real photon at long distances of order of $O(1/\sqrt{q^2})$.
Applying the LCSR approach~\cite{Kho99},
one effectively accounts for the long-distance effects
of a real photon,
using the quark-hadron duality in the vector channel
and a dispersion relation in $q^2$.

\begin{figure}[b]
 $$\includegraphics[width=0.9\textwidth]{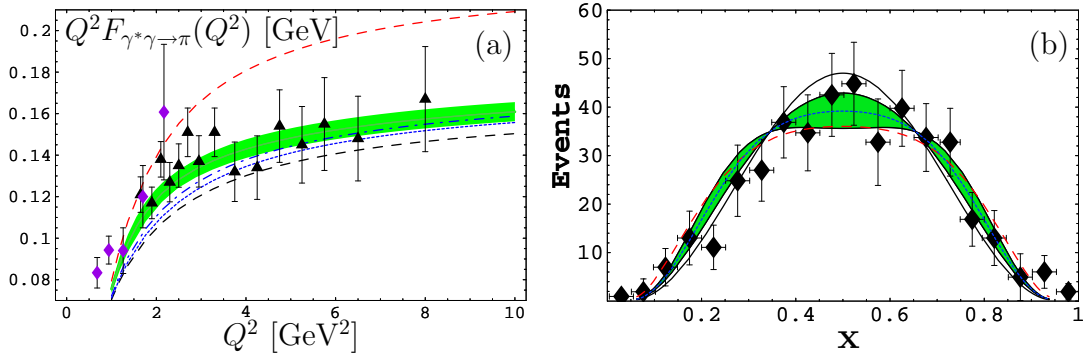}$$\vspace*{-9mm}

   \caption{\label{fig:CLEO_FF}\footnotesize
    (a) $\gamma^*\gamma\to\pi$ Transition form factor in comparison
    with the CELLO ({\ding{117}})~\cite{CELLO91} and
    the CLEO ({\ding{115}})~\cite{CLEO98} data. For details see in the text.
    (b)  Comparison of the asymptotic DA (solid line), Chernyak--Zhitnitsky (CZ)
    DA (dashed line)~\cite{CZ82},
    and the  BMS ``bunch'' of pion DAs (strip) with the E791 data (\ding{117})~\cite{E79102}.}
\end{figure}
In our CLEO data analysis \cite{BMS02},
we also took into account the relation  between
the ``nonlocality" scale and the twist-4 magnitude
$\delta^2_\text{Tw-4} \approx \lambda_q^2/2$,
which was used to  re-estimate
$\delta^2_\text{Tw-4}= 0.19 \pm 0.02$ at $\lambda_q^2=0.4$ GeV$^2$.
To make our conclusions more precise,
we have adopted a 20\% uncertainty in the magnitude of the twist-4 contribution,
$\delta_\text{Tw-4}^2 = 0.19\pm0.04$~GeV$^2$,
and produced new $1\sigma$-, $2\sigma$- and $3\sigma$-contours
dictated by the CLEO data~\cite{BMS03}.
We concluded that even with a 20\% uncertainty in $\delta_\text{Tw-4}^2$,
the CZ DA is excluded \textbf{at least} at the \textbf{$4\sigma$}-level,
whereas the asymptotic DA -- at the \textbf{$3\sigma$}-level.
Our ``bunch'' is inside the $1\sigma$-region
and other nonperturbative models are near the 3$\sigma$-boundary.
We show in Fig.\,\ref{fig:CLEO_FF}a
the plot of $Q^2F_{\gamma^*\gamma\to\pi}(Q^2)$
for our ``bunch'' (shaded strip), CZ DA (upper dashed line), asymptotic DA (lower dashed line),
and two instanton-based models (dotted~\cite{PPRWG99} and dash-dotted~\cite{PR01} lines)
in comparison with the CELLO and CLEO data.
We see that the BMS ``bunch'' describes rather well all data for $Q^2\gtrsim1.5$ GeV$^2$.

\section{Diffractive Dijet Production}
The diffractive dijet production in $\pi+A$ collisions has been suggested as a tool
to extract the profile of the pion DA by Frankfurt et al. in 1993~\cite{FMS93}.
They argued that the jet distribution
with respect to the longitudinal momentum fraction
has to follow the quark momentum distribution in the pion
and hence provides a direct measurement of the pion DA.
As it was shown just recently in~\cite{BISS02} (see also~\cite{NSS01}),
this proportionality does not hold beyond the leading logarithms in energy.
Braun et al. found that the distribution in the longitudinal momentum fraction
of the jets for the non-factorizable contribution
is the same as for the factorizable contribution
with the asymptotic pion DA.
Using the convolution approach of Braun et al.
we estimated~\cite{BMS03} the distribution of jets in this experiment
for our ``bunch'' of pion DAs and show the results in comparison with
$\varphi^{\rm as}$ and $\varphi^{\rm CZ}$ in Fig.\,\ref{fig:CLEO_FF}b.
It is interesting to note that the corresponding $\chi^2$ values are:
as -- 12.56; CZ -- 14.15; BMS -- 10.96 (accounting for 18 data points).
The main conclusion from this comparison:
\textbf{all three DAs are compatible with the E791 data}.
Hence, this experiment cannot serve as a safe profile indicator.\vspace*{2mm}

\begin{figure}[b]
 $$\includegraphics[width=0.9\textwidth]{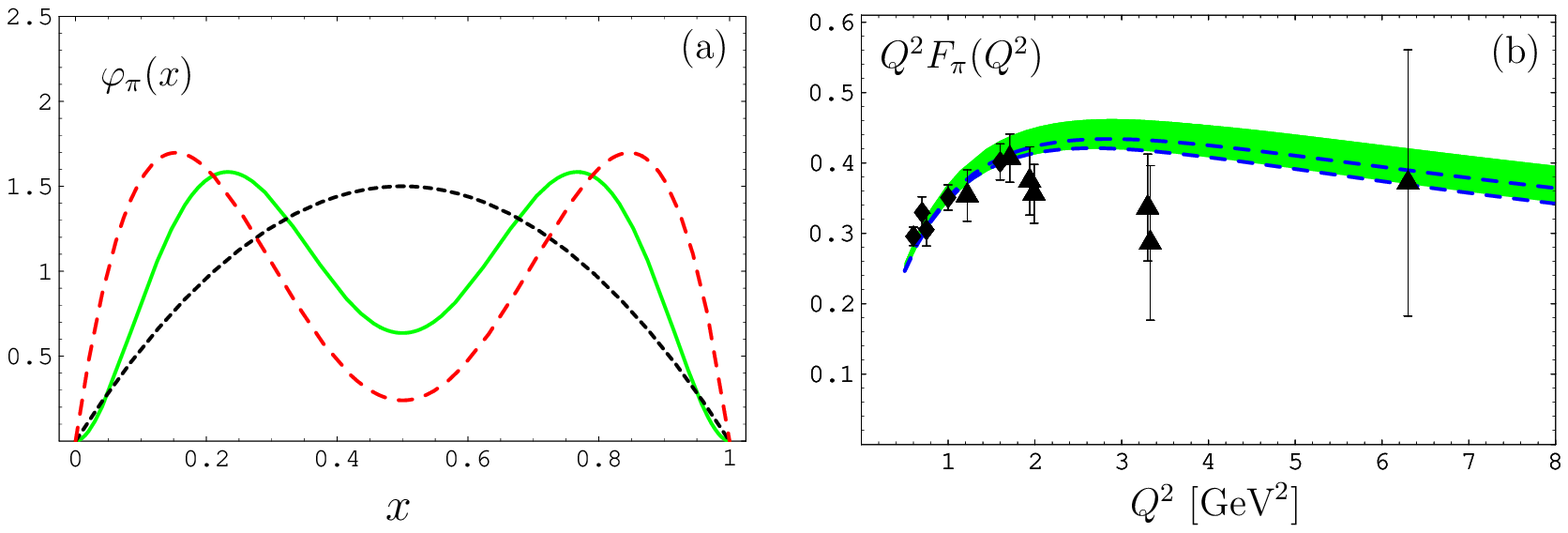}$$
 \vspace*{-8mm}

 \caption{\label{fig:pidatasum}\footnotesize
     (a) Comparison of $\varphi^{\rm as}$, 
         (dotted line), $\varphi^{\rm CZ}$ 
         (dashed line), and $\varphi^{\rm BMS}$ 
         (solid line).
     (b) Pion electromagnetic form factor in comparison
         with the JLab ({\ding{117}})~\protect\cite{JLab00}
         and Bebek et al. ({\ding{115}})~\protect\cite{FFPI73} data.
         Predictions based on the BMS ``bunch'' of pion DAs (green strip)
         including the NLC QCD sum-rule uncertainties and those due to
         scale-setting ambiguities at the NLO level.
         The region between the dashed lines denotes the area accessible to
         $\varphi^{\rm as}$.
         }
\end{figure}

Let us say a few words about similarities and differences
between the CZ and BMS DAs.
Both are two-humped,
but the CZ DA is strongly end-point enhanced,
whereas the BMS DA is end-point suppressed,
as is well illustrated in Fig.\,\ref{fig:pidatasum}a. 
And the reason for this behavior is physically evident:
nonlocal quark condensate reduces pion DA in the small $x$ region
and enhances in the vicinity of the point $x\simeq0.2$.
In order to keep the norm equal to unity,
it is forced to have in the central region some reduction as well.

\section{Pion electromagnetic form factor}
How well is the BMS bunch in comparison with the JLab data on the pion form factor?
We have calculated the pion form factor in analytic NLO pQCD~\cite{BPSS04}
\begin{equation}
  F_{\pi}(Q^{2};\mu_\text{R}^{2})
  =  F_{\pi}^\text{LD}(Q^{2})
  +  \left(\frac{Q^2}{2s_0^\text{2-loop}+Q^2}\right)^2
     F_{\pi}^\text{Fact}(Q^2;\mu_\text{R}^{2})\,,
 \label{eq:Q2Pff}
\end{equation}
taking into account the soft part $F_{\pi}^\text{LD}(Q^{2})$
via the local duality approach~\cite{NR82},
based on perturbative spectral density
$\rho(s_1,s_2,Q^2)$~\cite{IS82,NR82},
and correcting the factorized contribution $F_{\pi}^\text{Fact}$
via a power-behaved pre-factor
(with $s_0^\text{2-loop}\approx0.6$~GeV$^2$)
in order to respect the Ward identity
at $Q^2=0$ while preserving its high-$Q^2$ asymptotics.

In our analysis $F_{\pi}^\text{Fact}(Q^2;\mu_\text{R}^{2})$
has been computed to NLO~\cite{DR81,MNP99a}
using Analytic Perturbation Theory~\cite{SS97,DVS0012,SSK99}
and trading the running coupling and its powers for analytic expressions
in a non-power series expansion,
i.e.,
\begin{equation}
 \left[F_{\pi}^\text{Fact}(Q^2; \mu_\text{R}^{2})\right]_\text{MaxAn}
  \ =\ \bar{\alpha}_\text{s}^{(2)}(\mu_\text{R}^{2})\, {\cal F}_{\pi}^\text{LO}(Q^2)
   + \frac{1}{\pi}\,
      {\cal A}_{2}^{(2)}(\mu_\text{R}^{2})\,
       {\cal F}_{\pi}^\text{NLO}(Q^2;\mu_\text{R}^{2})\,,
\label{eq:pffMaxAn}
\end{equation}
with $\bar{\alpha}_\text{s}^{(2)}$ and ${\cal A}_{2}^{(2)}(\mu_\text{R}^{2})$
being the 2-loop analytic images of $\alpha_\text{s}(Q^2)$
and $\left(\alpha_\text{s}(Q^2)\right)^2$, respectively \cite{BPSS04},
whereas
${\cal F}_{\pi}^\text{LO}(Q^2)$ and
${\cal F}_{\pi}^\text{NLO}(Q^2;\mu_\text{R}^{2})$
are the LO and NLO parts of the factorized form factor.
This procedure with the analytic running coupling and the analytic versions of its powers
gives us a practical independence of the scheme/scale setting
and provides results in rather good agreement with the
experimental data \cite{FFPI73,JLab00}.
Indeed, the form-factor predictions are only slightly larger
than those resulting from the calculation with the asymptotic DA
(see Fig.\,\ref{fig:pidatasum}b).

\section{Conclusions}
The QCD sum rule method with nonlocal condensates
produces a ``bunch'' of admissible pion DAs
for each {$\lambda_q^2$} value.
Comparing these results
with the CLEO constraints,
obtained in the LCSR analysis of the $\gamma^*\gamma\to\pi$-transition
form factor,
clearly fixes the value of QCD vacuum nonlocality to
$\lambda_q^2=0.4$~GeV$^2$.
The corresponding ``bunch'' of pion DAs agrees well with both
the E791 data on diffractive dijet production and with
the JLab F(pi) data on the pion electromagnetic form factor.
Analytic perturbation theory with a non-power NLO contribution
for the pion form factor diminishes scale-setting ambiguities
already at the NLO level.

\begin{theacknowledgments}
This work was supported in part by the Deutsche Forschungsgemeinschaft
(projects 436KRO113/6/0-1 and 436RUS113/752/0-1),
the Heisenberg-Landau Programme,
and the Russian Foundation for Fundamental Research
(grants 03-02-16816 and 03-02-04022).
I am indebted to the organizers of the Conference
for financial support.
\end{theacknowledgments}


\end{document}